\documentclass[conference]{IEEEtran}
\IEEEoverridecommandlockouts
\usepackage{cite}
\usepackage{algorithm}  
\usepackage{algpseudocode} 
\usepackage{colortbl}
\definecolor{lightgray}{gray}{0.7}
\usepackage{amsmath,amssymb,amsfonts}
\usepackage{bbm}
\usepackage{mathrsfs}
\usepackage{dutchcal}
\usepackage{graphicx}
\usepackage{textcomp}
\usepackage{xcolor}
\usepackage{threeparttable}

\def\BibTeX{{\rm B\kern-.05em{\sc i\kern-.025em b}\kern-.08em
    T\kern-.1667em\lower.7ex\hbox{E}\kern-.125emX}}
\begin{document}

\title{Enhanced LSTM-based Service Decomposition for Mobile Augmented Reality}

\author{\IEEEauthorblockN{ 
Zhaohui Huang}
\IEEEauthorblockA{\textit{Department of Engineering} \\
\textit{King's College London}\\
London, Strand, WC2R 2LS, UK \\
zhaohui.huang@kcl.ac.uk}
\and
\IEEEauthorblockN{Vasilis Friderikos}
\IEEEauthorblockA{\textit{Department of Engineering} \\
\textit{King's College London}\\
London, Strand, WC2R 2LS, UK \\
vasilis.friderikos@kcl.ac.uk}

}
\maketitle

\begin{abstract}
Undoubtedly, Mobile Augmented Reality (MAR) applications for 5G and Beyond wireless networks are witnessing a notable attention recently. However, they require significant computational and storage resources at the end device and/or the network via Edge Cloud (EC) support. In this work, a MAR service is considered under the lenses of microservices where MAR service components can be decomposed and anchored at different locations ranging from the end device to different ECs in order to optimize the overall service and network efficiency. To this end, we propose a mobility aware MAR service decomposition using a Long Short Term Memory (LSTM) deep neural network to provide efficient pro-active decision making in real-time. More specifically, the LSTM deep neural network is trained with optimal solutions derived from a mathematical programming formulation in an offline manner. Then, decision making at the inference stage is used to optimize service decomposition of MAR services. A wide set of numerical investigations reveal that the mobility aware LSTM deep neural network manage to outperform recently proposed schemes in terms of both decision making quality as well as computational time.
\end{abstract}
\setlength{\parskip}{0em}

\section{Introduction}
\begin{figure}[htb]
	\centering
	\includegraphics[width=0.49
	\textwidth]{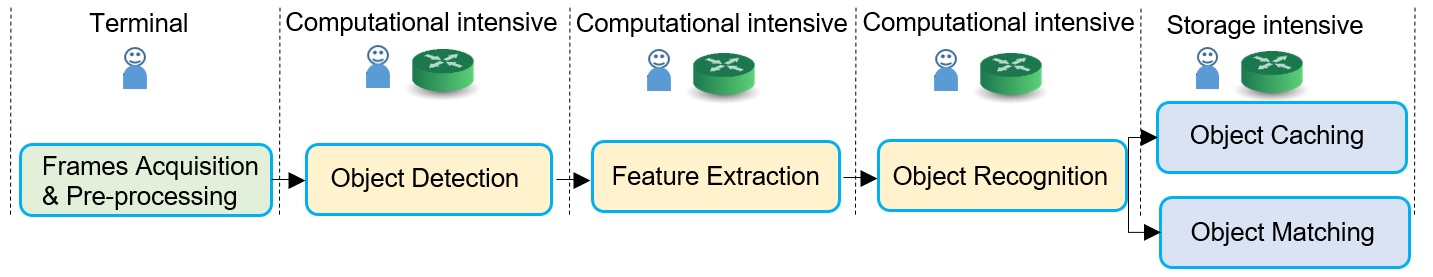}
	\caption{The flow of the different nominal mobile AR functions, their characteristics and the potential location where those functions can run (terminal and/or at the EC).}
	\label{fig:MAR functions}
\end{figure}

\IEEEPARstart{D}{ifferent} from traditional applications, Mobile augmented reality (MAR) enhance virtual experiences through elevating capabilities of mobile devices and hence brings about significant flexibility and accuracy in amalgamating Augmented Reality(AR) objects with the physical world. However, it has naturally a higher degree of requirements in computing and caching resources, especially when rendering 3-dimensional (3D) AR objects\cite{li2020rendering}. To this end, Edge clouds (ECs) assistance allows the deployment of service at the edge of the network and hence is widely embedded in the design for orchestrating and offloading MAR services. In  \cite{huang2021proactive}, we have outlined how a MAR application can be decomposed into a series of granular micro-services together with an optimization framework to provide optimal pro-active mobility-aware decision making in terms of EC assignment. As illustrated in Fig. \eqref{fig:MAR functions}, the captured video of a MAR application is preprocessed at the terminal and then frames with AR objects are transmitted for detection, extraction and recognition. The local cache is searched to find if there is a match. Finally, the matched results are transmitted back for presentation. According to their features, these MAR functions are categorized into two types: computational intensive ones that require significant CPU resources and storage intensive ones that require significant cache resources. Clearly, they have a predefined order when working as a service chain.  

\begin{figure}[htb]
	\centering
	\includegraphics[width=0.5\textwidth]{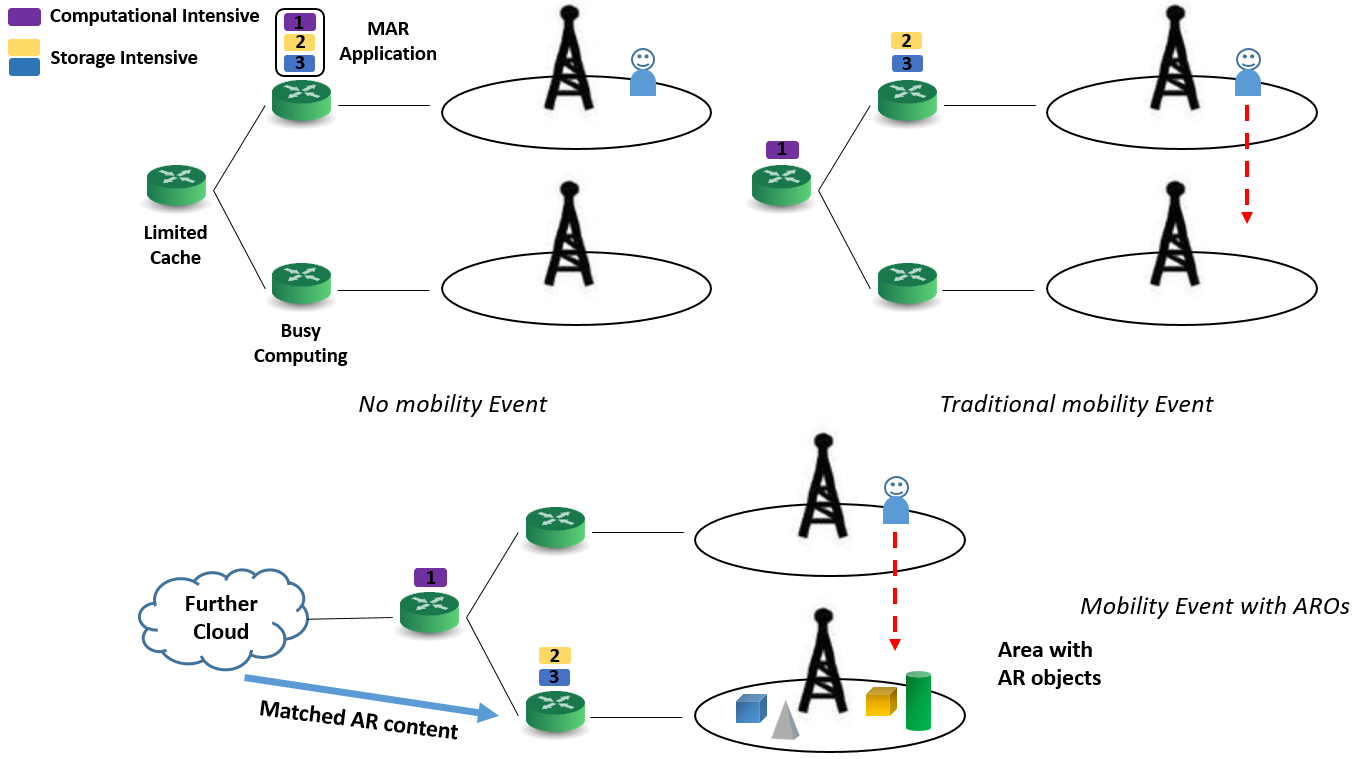}
	\caption{Illustrative toy examples of the effect of mobility on pro active resource allocation of MAR functions.  differences between applying a complete MAR application and its decomposed functions in a no mibility event, traditional mobility event and mobility event with AROs }
	\label{fig:toy}
\end{figure}

The benefit of applying such decomposition can be further revealed by a toy example shown in Fig. \eqref{fig:toy}. When there is no mobility like the first case showing in Fig. \eqref{fig:toy}, it could be possible to deploy the  MAR application at a single server. However, it is clear that, without decomposition, the complete MAR application uses significant levels of CPU and cache memory resources. On the other hand, decomposition allows a more flexible allocation but the proactive caching is still not necessary here because there is no target objects distributed in the area. When bringing in both AR objects and the user mobility, an isolated module could be designed to provide predictions for proactive resource allocation according to target AR objects and the user's possible future destination. Thus, the given prediction not only contains where to set functions and send requests, but also indicates what AR content should be allocated to the target server. In the Optim scheme from \cite{huang2021proactive}, constructing and solving a complex mixed integer linear problem is quite time-consuming and cannot respond to network changes in time. Therefore, the previous scheme should be improved to take into account both high-quality solutions and computational efficiency.

Leveraging machine learning (ML) techniques on the network, such as neural networks or reinforcement learning, are considered to have the potential to solve problems in this scope area. Till now, a series of several ML-based solutions have been proposed with improved prediction capabilities, flow path management and resource allocation features \cite{bianchini2020toward}. However, the best way to address these issues with ML techniques still remains unclear. In \cite{nath2019ptc}, an orchestration framework entitled Pick-Test-Choose (PTC) is proposed. It contains an iterative reinforcement machine learning algorithm based on Bayesian optimization. Workload characteristics and resource availability are monitored and predicted to decide a suitable fog device location. Before making a final decision, this choice is tested to see if it satisfies the expectation and constraints. Thus, their design becomes more adaptive to the network environment. ML techniques are integrated as isolated modules in \cite{jalodia2019deep}. To provide resource predictions in dynamic network function virtualization (NFV) environments, a model applying deep neural network (DNN) is proposed. Their learned policy is reinforced according to asynchronous observations provided by multiple agents in the network. Such algorithm working in deep reinforcement learning (DRL) agent can provide automated dynamic scaling solutions for resource allocation with advantages on handling a large number of outcomes and impact over time\cite{jalodia2019deep}. Related research in this area reveal the advantages of integrating ML techniques especially on solving resource allocation problems in different types of networks. Thus, a well-known ML technique, long short term memory (LSTM), is applied in this paper to learn from optimal solutions and provide reliable predictions.

LSTM is widely accepted as an enhanced recurrent neural network due to its ability in memorizing significant information and forgetting unnecessary inferences to overcome potential gradient vanishing and gradient exploding problems \cite{hochreiter1997long}\cite{zuo2019learning}. In a closely related work from \cite{zuo2019learning}, the traffic path planning problem is also focused. A sequence to sequence model is constructed to learn implicit forwarding paths from empirical traffic records. In their seq2seq model, LSTM cells are applied as basic instances of encoder and decoder to bridge the gap between input and output. An attention scheme is created to ensure a reasonable linked order for all elements in output sequences by scoring their relevance between source and target. To avoid the local optimum problem in the long run and boost the overall performance of the model, they propose beam search that widens the searching range and verifies the connectivity of forwarding paths\cite{zuo2019learning}. Similarly, a LSTM-based model is also applied in this paper to train for high-quality path predictions in an edge cloud supported network. However, as pointed out by our previous work, the execution sequence of two main types of functions is pre-defined during the decomposition of MAR services. Therefore, the correct order of the path could be figured out whenever the assignment is provided. Another significant difference is our model directly learns from optimal solutions generated by the algorithm proposed in our previous work, which weakens the advantage of beam search in our case. However, the validation of learning results still remains as a problem. Thus, a feasibility check stage is added to ensure the predictions can still satisfy constraints and fit the original network environment. 

As already eluded in this work optimal decision making is utilized to train the deep neural network. The mathematical programming formulation that is used to provide optimal decision making is detailed in \cite{huang2021proactive};in that case, an edge cloud resource allocation optimization framework is proposed to proactively allocate resources and to satisfy related requirements of MAR applications. The user mobility is considered explicitly and the overall delay is minimized as  multi-objective optimization problem. 
Although the optimal solution is desirable however such a framework cannot be used to provide real-time decision making since solving a mixed integer mathematical program suffers from the curse of dimensionality which means that requires significant amount of time to provide a solution. Therefore, in this paper, an LSTM-based approach is explored where the optimal solutions are used to train the deep neural network in an offline manner. 
\section{System Model}
A set $\mathbb{M}=\{1,2,...,M\}$ is defined to represent the available Edge Clouds (ECs) in the network. Multiple mobile devices are assumed to create MAR service requests $r\in R$. With $\eta$ and $\varrho$ we define computational intensive and storage intensive  MAR functionalities respectively  \cite{huang2021proactive}. The allowed destinations for a user are limited to adjacent access routers. Thus, The probability of a user moving from the initial location to an adjacent server can be defined as $p_{rij} \in [0,1]$, where adjacent servers ${i,j} \subset \mathbb{M}$. The location of executing one functionality for a request can be denoted as $L_{\eta r}\in\mathbb{M}$ or $L_{\varrho r}\in\mathbb{M}$. Also, in a typical MAR service, each request requires the execution of two functionalities in a pre-defined order \cite{huang2021proactive}. Thus, the assignment of one request can be arranged according to the route of its flow. The route matrix can be denoted as $RT_r=[s_r,L_{\eta r},L_{\varrho r},d_r]$. This matrix can be provided by the optimal scheme and consist a route assignment set $\mathbb{RT}$ \cite{huang2021proactive}.
In LSTM training, requests' initial locations and their destinations can be fed as inputs and decisions of where to anchor MAR functionalities can be treated as the output. Thus, we further split the route matrix into $X_r=[s_r,d_r]\in\mathbb{X}$ and $Y_r=[L_{\eta r},L_{\varrho r}]\in\mathbb{Y}$. The major part of the route set $\mathbb{RT}$ is used as input and fed into the LSTM-based model for training. The rest of this set remains as a testing set, Therefore, we denote with $\mathbbm{XTrain} ,\mathbbm{XTest} \subset\mathbb{X}$ and $\mathbbm{YTrain},\mathbbm{YTest}\subset\mathbb{Y}$ to differentiate training and testing sets. The output is the predictions made by the trained model and hence denoted as $\mathbbm{YPred}$. 

At first, the optimal scheme is called to track the MAR requests in the network. The mathematical programming formulation (which is a Mixed Integer Linear Problem - MILP) captures both the computational delay in two types of functions and the EC routing delay from the terminal to the target servers and provides optimal decision making for anchoring each MAR service functionalities to different ECs \cite{huang2021proactive}. 

\begin{figure}[htb]
	\centering
	\includegraphics[width=0.42\textwidth]{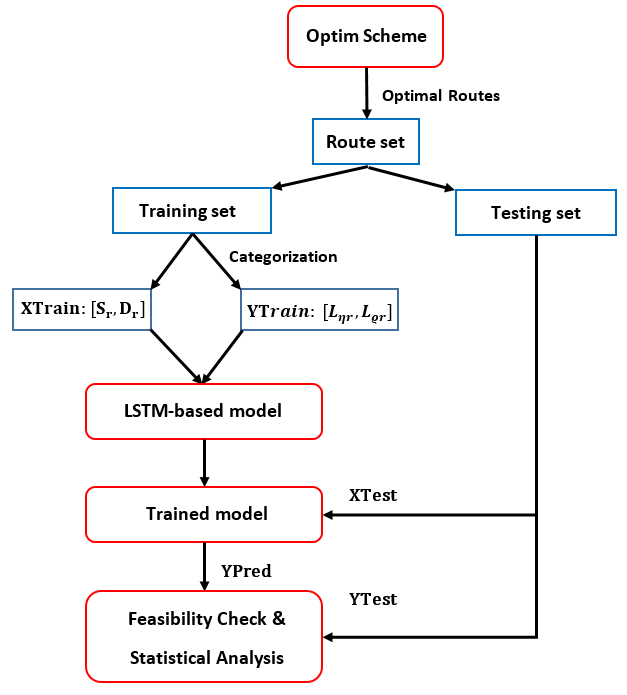}
	\caption{Working process}
	\label{fig:in/ouput process}
\end{figure}

The process from obtaining assignments to providing predictions is shown by Fig. \eqref{fig:in/ouput process}. As mentioned, the Optimal assignments are generated by the MILP scheme are grouped and reshaped into a route matrix and then separated for training and testing. It is necessary to point out that $\mathbbm{YTrain}$ is further adapted to categorical type to enable the proposed  LSTM-based model to better complete the classification work. Since execution locations for functionalities can only be selected from the set $\mathbb{M}$, all possible different results can be calculated as selecting any two from $M$ elements with order, which is $M^2$. For simplicity, each assignment can be given an index to show which type it belongs to. Thus, $Y_r$ can now be represented by $Ind_r\subset\{1,...,M^2\}$. The LSTM-based model will then try to classify and figure out the relation between two input sets. After training, the model will provide its predictions according to the input training set. The predictions will be transferred back into locations and sent for Feasibility Check. In Feasibility Check, the predictions are put back into the original network environment and check whether they still satisfy all constraints such as cache size and EC capacity limit. If an EC is overloaded, the extra predicted request will be sent to an available neighbor EC as a backup choice. However, if both ECs are occupied, then the request is sent to a cloud deeper in the network and, as a result, an extra cost penalty is triggered. In order to shed light into the quality of the decision making, the assignments are  compared with the optimal assignments to evaluate the quality of the proposed mechanism.

\begin{figure}[htb]
	\centering
	\includegraphics[width=0.42\textwidth]{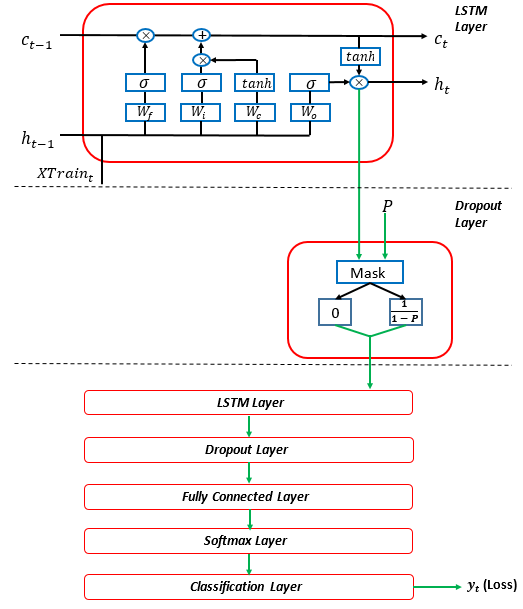}
	\caption{Layered structure of LSTM-based model (at time step $t$)}
	\label{fig:design}
\end{figure}

The architecture of the LSTM-based model is shown in Fig. \eqref{fig:design}. It is clear that the LSTM layer follows the classic definition and its state changing and gate controlling still follows the original formula. The aim is to explore the potential of using a nominal LSTM network for service decomposition. However, note that the number of the total time steps equals the dimensions of the input matrix $\mathbbm{XTrain}$. It is followed by a dropout layer to avoid over-fitting by set a random number of elements to $0$ ($rand(size(X))<P$, $X$ is a layer input and $P$ is a self-defined probability) and scaling other elements by $\frac{1}{1-P}$ \cite{srivastava2014dropout}\cite{krizhevsky2017imagenet}. In this design, two LSTM layers with their dropout layers are created to enhance the overall performance. Then in the fully connected layer, the sent result is multiplied by a weight matrix and added by a bias vector \cite{glorot2010understanding}\cite{he2015delving}. This layer combines features and tries to identify a larger pattern \cite{glorot2010understanding}\cite{he2015delving}. In the softmax layer, Without particularly specified, a logistic sigmoid function will work as the output unit activation function \cite{bishop2006pattern}. Finally, the classification layer takes values from the previous softmax function and calculates the cross entropy loss at the time step $t$, which is $y_t$. In the following time steps, this process will repeat until finishing the whole training.

\section{Numerical Investigations}
Hereafter, the effectiveness of the proposed LSTM-based scheme is investigated via a wide-set of simulations and compared with baseline schemes. 

\begin{figure}[htb]
	\centering
	\includegraphics[width=0.45\textwidth]{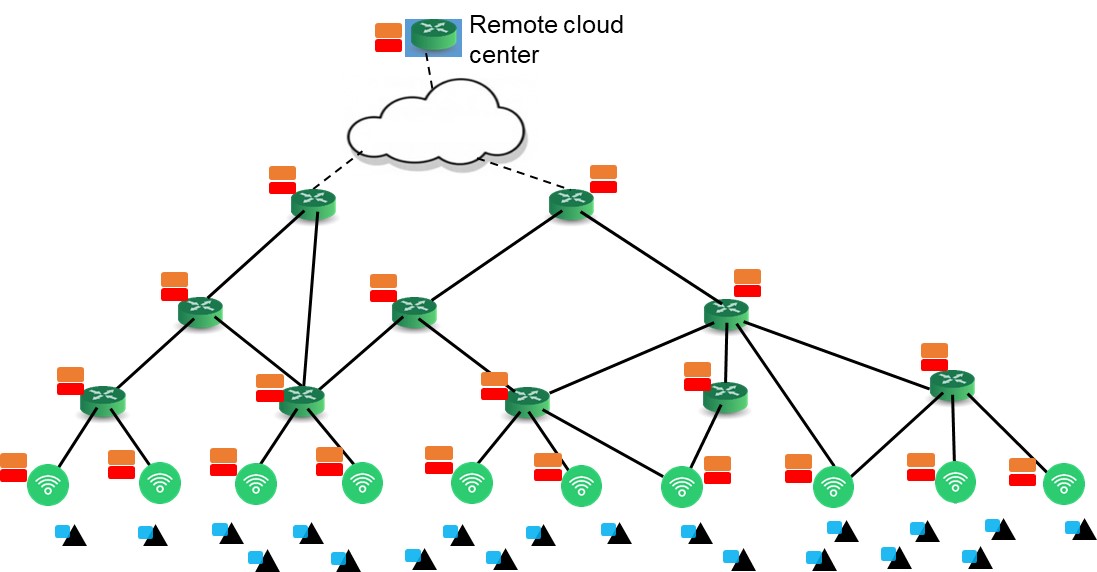}
	\caption{Typical tree-like designed network topology, with dense EC support, however a randomized small subset of those are assumed active in the simulations. \cite{huang2021proactive}}
	\label{fig:topology}
\end{figure}

The set of ECs and the network topology is shown in Fig. \eqref{fig:topology}. We assume 20 ECs in total with 6 ECs being activated and up to 40 requests are sent by MAR devices in each simulation. The remaining available resources of an EC is assumed as a CPU with 4 to 8 cores and 16GB memory\cite{huang2021proactive}. Each request requires a single free unit for each service function (for example a VM) and each target ARO is around 0.12MB to 2MB \cite{liu2018edge}. In each EC, up to 14 units are assumed inside as its capacity and take up CPU resources equally \cite{huang2021proactive}. As eluded previously, users move between adjacent access routers (ARs) from the initial location. The probability of mobility of a user $r$ between ARs $i,j$ is denoted as $p_{rij}$  ($\sum_{j \in \mathbb{M}} p_{rij} \leq 1 $). Optimal solutions are calculated based on the model in \cite{huang2021proactive} and 5 simulations are repeated to provide sufficient solutions to form a suitable size of the assignment set; 90\% of this set is fed into the model as a training set while the rest 10\% is used for testing. In the LSTM layer, the initial learning rate is set to 0.005 and the maximum number of epochs is 160. 

Key parameters used are described in Table \eqref{tab:parameters}. The probability in the dropout layer is set to 5\% to avoid over-fitting. Several different baseline schemes are implemented as well for comparison. The optimal solution scheme (Optim)  provides the optimal solutions by solving the underlying MILP problems \cite{huang2021proactive}. The FACT scheme also considers the balance between frame size and service delay and tries to find optimal solutions without taking the user's mobility effect into account\cite{liu2018edge}. The random selection scheme (RandS) selects the target servers randomly whilst the closest-first scheme (CFS) tends to choose the nearest available one to the user's initial location \cite{tocze2019orch}. The utilization scheme (UTIL) takes the least loaded neighbor EC as a backup if the nearest one excesses an availability threshold status (for example, more than 80\% of total VMs are occupied)\cite{sonmez2019fuzzy}. Except for the optimal scheme, all other schemes would need to be modified to satisfy a set of constraints so that their comparison could be deemed as fair. 
\begin{table}[!htbp]
    \centering
    \caption{Key Parameters of LSTM-based network}
    \begin{tabular}{c|c|l}
         \hline
         \textbf{Parameter}&\textbf{Value}&\textbf{Description}\\
         \hline
         numEC&$6$& the number of activated ECs \\
         \hline
         $p_{rij}$&$[0,1]$& a user's moving probability from the initial\\&&location to an adjacent server\\
         \hline
         XTrain,XTest&$135$& initial location \& destination pairs from Optim\\ 
         \hline
         YTrain,YTest&$15$& pairs of assignment from Optim\\
         \hline
         numRes&36& types of responses\\
         \hline
         MaxEpochs&160& the number of allowed maximum iterations\\
         \hline
         numHidden&80& the number of hidden units\\
         \hline
         DropRate&0.05& the probability of dropping out \\
         \hline
         InitialLearn&0.005& the initial learning rate\\
         \hline
    \end{tabular}
    \label{tab:parameters}
\end{table}

\begin{figure}[htb]
	\centering
	\includegraphics[width=0.42\textwidth]{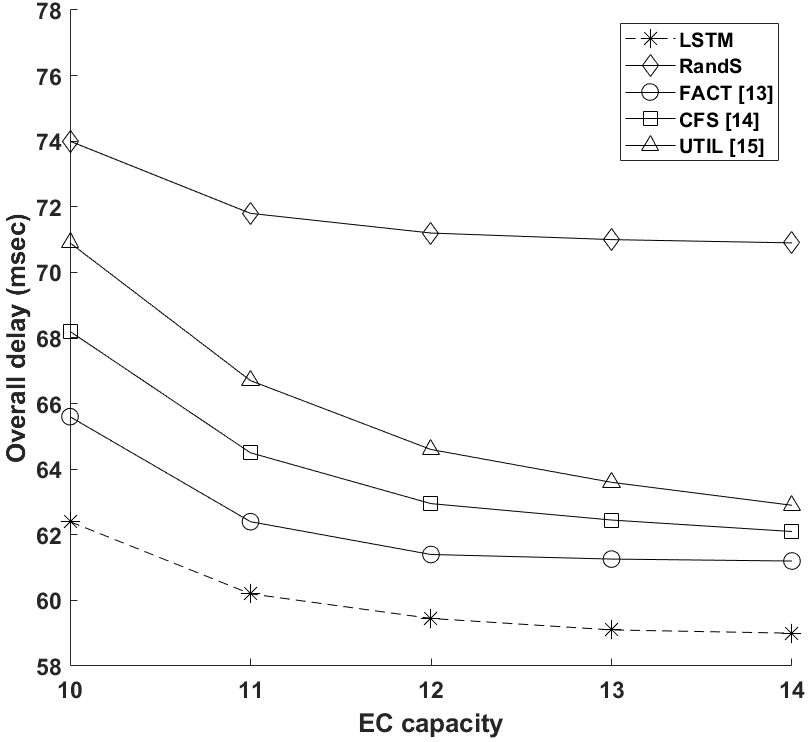}
	\caption{Average service delay of different schemes with different EC capacities (6 ECs and 30 requests)}
	\label{fig:capacity}
\end{figure}

\begin{figure}[htb]
	\centering
	\includegraphics[width=0.42\textwidth]{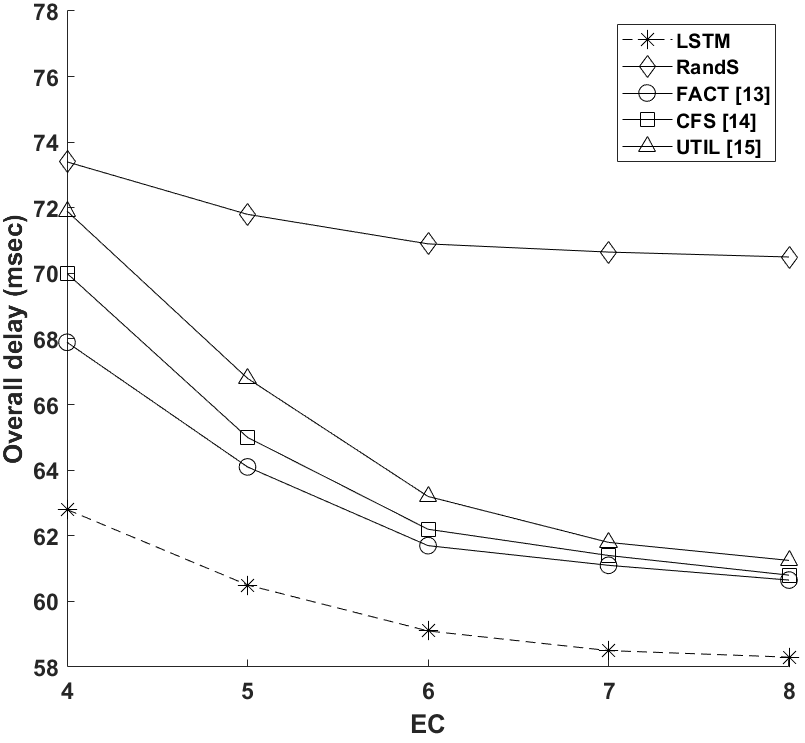}
	\caption{Average service delay of different schemes with different numbers of EC (30 requests and Capacity is 14)}
	\label{fig:EC}
\end{figure}

\begin{figure}[htb]
	\centering
	\includegraphics[width=0.42\textwidth]{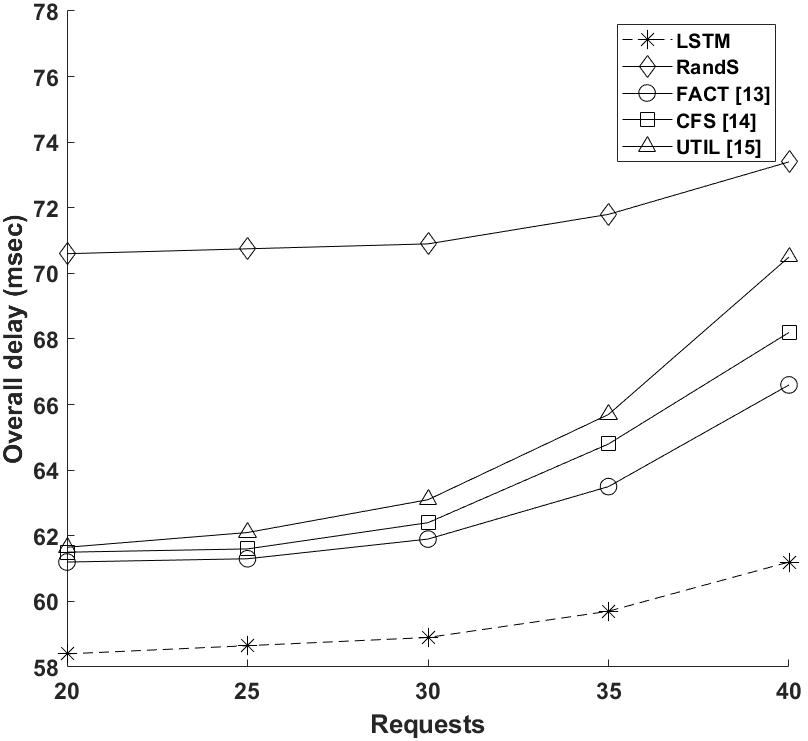}
	\caption{Average service delay of different schemes with different numbers of requests (6 ECs and Capacity is 14)}
	\label{fig:Req}
\end{figure}

Fig. \eqref{fig:capacity} shows the difference in performance between the LSTM-based model and other baseline schemes when the EC capacity increases from 10 to 14 units. It is clear that the proposed scheme can provide better decision making than all other schemes. Its advantage is more obvious in a congested network. Most greedy schemes share a common tendency to execute decomposed MAR services on a few ECs. The flexibility of decomposition is neglected and hence these greedy schemes will suffer from a serious penalty because of the narrow space and limited resources. The Fact scheme is better than them but still neglects the mobility effect. LSTM-based model, on the other hand, is not seriously affected by congestion and only keeps focus on learning from the Optim scheme which considers the mobility and makes full use of decomposition. Similarly, their performances under other settings like different number of ECs and requests are shown by Fig. \eqref{fig:EC} and \eqref{fig:Req}. Clearly, the proposed scheme is less sensitive than other baseline schemes and hence is more reliable in a congested network. According to Table \eqref{tab:no_mob}, when there is no mobility and the network is not congested, the advantage of the Optim scheme is not obvious. The Optim scheme shares the same performance as the FACT scheme and is also close to the CFS scheme. The LSTM-based scheme still maintains its gap to the Optim scheme. Compared with other schemes, its performance is still acceptable in this case.     
\begin{table}[ht]
\caption{Delay for all schemes with no mobility (6 ECs, 30 requests and Capacity is 14)}
    \centering
    \begin{tabular}{c|c|c|c|c|c}
    \hline
         \textbf{LSTM}&\textbf{Optim}&\textbf{FACT}&\textbf{CFS}&\textbf{RandS}&\textbf{UTIL}\\
         \hline
         48.4&47.2&47.2&47.7&56.0&50.8 \\
         \hline
    \end{tabular}
    \label{tab:no_mob}
\end{table}

\begin{figure}[htb]
	\centering
	\includegraphics[width=0.45\textwidth]{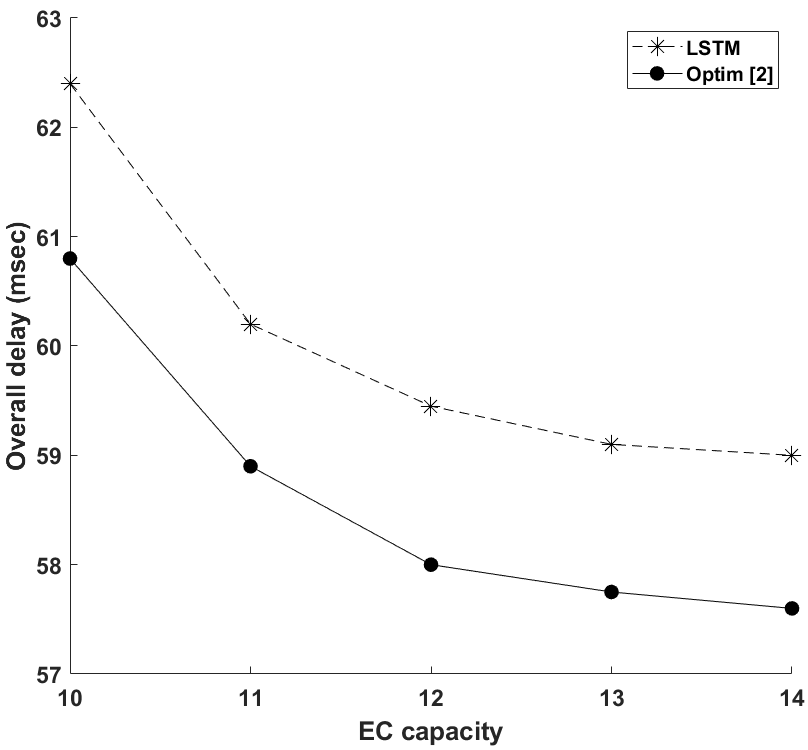}
	\caption{Difference in performance between LSTM and Optim}
	\label{fig:zoom}
\end{figure}

Since the difference between the LSTM-based model and the Optim scheme is quite narrow, a zoom version of EC capacity focusing on these two schemes is provided as Fig. \eqref{fig:zoom}. LSTM-based model is around 1.2 to 1.7 msec worse than the Optim scheme.    

In simulations, when comparing predictions and optimal solutions, we also have $rmse< 1$ (root mean square error), $\delta<11\%$ (relative error) and $r^2>0.88$ (determination coefficient) in most cases, which indicate the predictions can be deemed as reliable and highly competitive to the optimal solutions. According to Formula \eqref{formula:rmse} and \eqref{formula:re}, small values of $rmse$ and $\delta$ indicates that the proposed scheme generates a very similar version to the Optim's solution. The comparison of $rmse$ values of all schemes is shown by Table \eqref{tab:rmse}. 
We note that the indexes have a location information, i.e., access routers which are topologically close have also adjacent indices.
In addition, the user mobility is limited to adjacent EC within each period which is a valid assumption for active MAR sessions. Thus, high similarity to the optimal solutions usually indicates that the corresponding scheme has the ability to provide high-quality solutions as well. The value of $r^2$ is close to 1 and hence the proposed model shows a high quality fitting ability in this problem. Clearly, the quality will deteriorate when the output solution lead to an infeasible allocation. In that case, requests will have to be served by a cloud server which will entail some penalty. Experimentation showed that this was rarely the case during the training set but tends to appear in the testing set.
Table \eqref{tab:Acc} shows the variation of training and validation accuracy. 
Through the training, the proposed scheme finally memorizes the categories of given assignments and matches the training set. The validation accuracy increases with the training accuracy and finally maintains stable at around 94.6\%.  
\begin{equation}
\begin{aligned}
rmse=\sqrt{\frac{\Sigma_1^n(YTest_i-YPred_i)^2}{n}}  
\end{aligned}
\label{formula:rmse}
\end{equation}
\begin{equation}
\begin{aligned}
\delta={\frac{\Sigma_1^n|\frac{YTest_i}{YPred_i}-1|}{n}}\times 100\%
\end{aligned}
\label{formula:re}
\end{equation}
\begin{table}[htb]
    \centering
    \caption{RMSE Values of All Schemes}
    \begin{tabular}{c|c|c|c|c|c}
    \hline
         \textbf{Algorithm}&LSTM&RandS&CFS&FACT&UTIL  \\
         \hline
         \textbf{RMSE}&0.9&6.8&1.9&1.5&3.4 \\
         \hline
    \end{tabular}
    \label{tab:rmse}
\end{table}
\begin{table}[htb]
    \centering
    \caption{Variation of Training and Validation Accuracy}
    \begin{tabular}{c|c|c|c|c|c}
    \hline
         \textbf{Iterations}&2&40&80&120&160  \\
         \hline
         \textbf{Training}&22.5&71.9&92.0&99.4&100.0\\
         \hline
         \textbf{Validation}&21.3&66.2&83.8&93.7&94.6 \\
         \hline
    \end{tabular}
    \label{tab:Acc}
\end{table}

\begin{figure}[htb]
	\centering
	\includegraphics[width=0.42\textwidth]{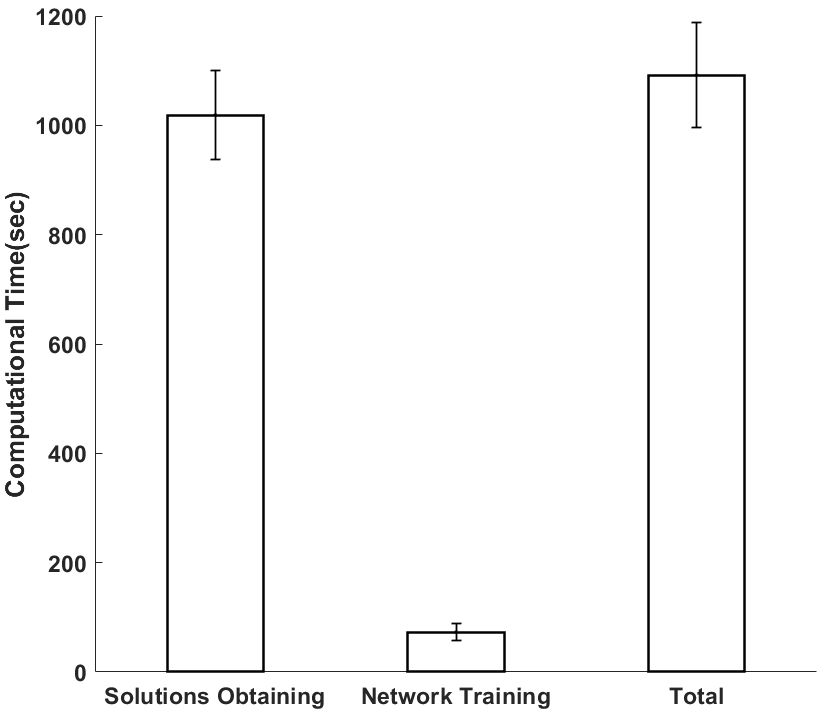}
	\caption{Average execution time of the training phase (6 ECs, 30 Requests and 14 units of Capacity)}
	\label{fig:PrepTime}
\end{figure}

Although obtaining optimal solutions and training the model based on the Optim scheme can be  time-consuming, the inference stage itself is highly efficient. Also, the time required to train the network dominated by the time to provide the optimal solutions as shown in Fig. \eqref{fig:PrepTime}.
The average processing time of the inference stage is compared with other algorithms in Table \eqref{tab:processing-time} (executed once). Considering the fact that optimal solutions obtaining and network training stage only execute offline, the inference stage has a more dominant effect. Other greedy schemes need to check and handle multiple constraints (e.g. capacity and cache size) repeatedly in iterations and lead to a longer processing time. In the Optim scheme, a complex MILP problem is constructed and solved to find optimal solutions and hence is the most time-consuming one. Thus, the inference stage of the proposed scheme is the most efficient one among all schemes. This is of a significant importance for network operation since it achieves better performance and takes much less  time to provide decision making than other algorithms.
\begin{table}[ht]
\caption{Average processing time of algorithms}
\begin{center}
\begin{threeparttable}
\begin{tabular}{c|c|c}
\hline
\textbf{Algorithm}&\textbf{Average Processing time(sec)}&\textbf{STD} \\
\hline
RandS&1.076&0.151\\
\hline
CFS&1.083&0.156\\
\hline
UTIL&1.091&0.155\\
\hline
LSTM&0.427&0.065\\
\hline
Optim \cite{huang2021proactive}&201.506&21.965\\
\hline
\end{tabular}
\label{tab:processing-time}
\begin{tablenotes}
\item \qquad*tested by PC, intel i7, 6500U, 2 cores 
\end{tablenotes}
\end{threeparttable}
\end{center}
\end{table}

\section{Conclusions}
Mobile Augmented Reality (MAR) applications are sensitive to the user mobility and service delay. Service decomposition together with edge clouds allow for a more flexible resource allocation to better tackle the mobility event with AR objects for MAR applications. This paper further explores an LSTM-based scheme to better respond to changes in the network. By learning from optimal solutions offline, the scheme can efficiently provide high-quality decision making. This is validated by a series of simulations showing that the proposed scheme outperforms previously proposed solutions. Future avenues of research will articulate on the use of advanced meta-heuristic algorithms to provide near-optimal decision making for training the deep neural network in large network instances where optimal decision making is not efficient.

\bibliographystyle{IEEEtran}
\bibliography{bibliography.bib}

\end{document}